\documentclass[10pt, conference]{IEEEtran}
\usepackage{graphicx}

\ifCLASSINFOpdf
\else
\fi
\begin{document}
%
\title{Impact of Directional Receiving Antennas on Wireless Networks}


\author{\IEEEauthorblockN{Jean-Marc Kelif and Olivier Simon}
\IEEEauthorblockA{Orange Labs\\
38-40 rue du G\'en\'eral Leclerc, 92130 Issy-Les-Moulineaux, France\\
\{jeanmarc.kelif, olivier.simon\}@orange.com}
}


%


\maketitle

\begin{abstract}
We are interested in high data rates internet access, by the mean of LTE based wireless networks. 
In the aim to improve performance of wireless networks, we propose an approach focused on the use of UE equipped by directional receiving antennas. Indeed, these antennas allow to mitigate the interference and to improve the link budget. Therefore, the Signal to Interference plus Noise Ratio (SINR) can be improved, and consequently the performance and quality of service (QoS), too. 
We establish the analytical expression of the SINR reached by a user with directional antenna, whatever its location. This expression shows that directional antennas allow an improvement of  the SINR, and to quantify it. 
We develop different scenarios to compare the use of directional antennas instead of omnidirectional ones. They allow to quantify the impact of directional antennas in terms of performance and QoS.

\end{abstract}


%

\section{Introduction}

With the development of internet services and the exploding traffic demand, one of the great challenges for telecommunications operators is to offer high data rates services wherever the user is. Indeed, the deployment of broadband internet access in rural areas is costly since it usually requires building or upgrading a large area wired access networks for very few people. A solution to decrease this cost consists in offering internet high data rates services by using wireless access. 
In this case, rural area users are connected to the network by using a radio access mean, i.e. a radio base station, connected to the internet, transmitting and receiving data. With similar data rates as ADSL, LTE is particularly interesting to be used for fixed rural internet access. 

Unfortunately, LTE is highly sensitive to interferences. The challenge consists in decreasing the interferences in order to increase the SINR and capacity.
Different solutions are possible to mitigate interferences and they are to a large extent complementary: Antenna parameters planning, Inter-Cell Interferences Coordination (ICIC), Coordinated multi-point (CoMP), High-order Multiple Input Multiple Output (MIMO), Advanced receivers such as Interference Rejection Combining (IRC) etc...
 Another solution is proposed in this paper: using terminals with directional antennas. This solution requires static position of the terminal and is obviously specific to fixed radio services. Directional antennas should allow minimizing the interferences and, at the same time, significantly improving the link budget, thus allowing large cells deployment and higher network capacity.

Inter-Cell Interferences Coordination (ICIC) \cite{Lee12} \cite{Kos13}  \cite{iichetnet} is another technique allowing the limitation of interferences impact. The efficiency of this technique depends on the 'coordinations' between cells that can be done.
Coordinated Multi-Point (CoMP) transmission techniques \cite{Zha12} \cite{Ngu11} allow to limit the interferences or have limited capability of interferences suppression. CoMP requires the transmitters to share channel-state information
(CSI).
Multiple Input Multiple Output (MIMO) techniques \cite{Che11} \cite{Sch11} allow to improve the SINR. This technique is based on the deployment of multiple transmitting antennas and mutiple receiving antennas.

\underline{Our Contribution}: In this article, we develop an analytical approach which allows to easily calculate the SINR achievable in a given area, by users equipped with directional receiving antennas or omnidirectional receiving antennas. It allows the calculation of the CDF (Cumulative Distributed Function) of the SINR of user equipments (UE). Since this CDF characterizes the performance and the QoS, it becomes easy to do an evaluation of the impact of the integration of directional receiving antennas on UE. 

The paper is organized as follows. In Section \ref{model}, we present the system model. In particular, we establish the analytical expression of the SINR of a UE equipped with a directional or an omnidirectional antenna.
In Section \ref{scenario}, the scenarios are described. Section \ref{Resultats} presents the results obtained. Section \ref{conclude} concludes the paper.

\section{System Model} \label{model}

We consider a wireless network composed of $S$ geographical sites, composed by 3 base stations (BS). Each BS covers a sectored cell. We focus our analysis on the downlink, in the context of an OFDMA based wireless network (e.g. WiMax, LTE).\\
Let us consider:
\begin{itemize}

\item ~${\mathcal S}=\{1,\ldots,S\}$ the set of geographic sites, uniformly and regularly distributed over the two-dimensional plane. 

\item ~${\mathcal N}=\{1,\ldots,N\}$ the set of BS, uniformly and regularly distributed over the two-dimensional plane. The BS are equipped by directional antennas (fig. \ref{reseauDir}): $N$= 3 $S$. 

\item $H$ sub-channels $h\in\mathcal{H}=\{1,\ldots,H\}$ where we denote $W$ the bandwidth of each sub-channel.

\item Each sub-channel consists in a fixed number of sub-carriers.

\item $P_{fh}^{(k)}(u)$ the transmitted power assigned by base station $k$ to sub-carrier~$f$ in sub-channel~$h$ towards user $u$. 
\item $g_{fh}^{(k)}(u)$ the propagation gain between transmitter~$k$ and user $u$ in sub-carrier~$f$ and sub-channel~$h$.
\end{itemize}

We assume that time is divided into slots. Each slot consists in a given sequence of OFDMA symbols.
Since the time is slotted, transmissions within each cell do not interfere one with each other. 
We assume that there is no interference between sub-carriers.
The total amount of power received by an UE $u$ connected to a BS $k_0$, on sub-carrier $f$ of sub-channel $h$ is given by the sum of : useful signal $P_{fh}^{(k_0)}(u) g_{fh}^{(k_0)}(u)$, interference due to the other transmitters $\sum\limits_{k\in\mathcal{N},k\neq k_0}P_{fh}^{(k)}(u) g_{fh}^{k}(u)$ and thermal noise $N_{th}$.\\
We consider the SINR $\gamma_{fh}(u)$ defined by:
\begin{equation} \label{SINR}
\gamma_{fh}(u)=\frac{P_{fh}^{(k_0)}(u) g_{fh}^{(k_0)}(u) }{
\sum\limits_{k\in\mathcal{N},k\neq k_0}P_{fh}^{(k)}(u) g_{fh}^{k}(u)  + N_{th}}
\end{equation}
as the criterion of radio quality. 

As we investigate the quality of service and performance issues of a network composed of omnidirectional receiving antennas and directional ones, the scenarios analyzed consider that all the subcarriers are allocated to UEs. Consequently, each sub-carrier $f$ of the sub-channel $h$ of any BS is used and can be an interferer of the ones of other BS. So we can drop the indexes $f$ and $h$.

\subsection{Propagation} \label{propagation}
Let us consider a path gain $g = Kr^{-\eta}$X, where  $K$ is a constant, $r$ is the distance between a transmitter $t$ and a receiver $u$,  and $\eta >$ 2 is the path loss exponent. The parameter X characterizes the shadowing expressed as a lognormal random variable.

Let us consider a user $M$ connected at the BS $i$, located at distance $r_i$ from it and angle $\theta_i$ with the direction of this antenna (Fig. \ref{dirantenna}). 

\subsubsection{Omnidirectional receivers} \label{antennaomni}

Equipped with an omnidirectional antenna, this UE receives a useful power $P_o$ expressed as  
\begin{equation}
P_o(r_i, \theta_i) = P_t Kr_i^{-\eta}G_T(\theta_i) X_i 
\end{equation}
where $P_t$ the transmitted power, $G_T(\theta_i)$ is the antenna gain of the transmitting antenna of the BS$_i$, and $X_i$ represents the shadowing. 

\subsubsection{Directional receivers} \label{antennadirect}
When this user is equipped by a directional antenna, he receives a power $P_d$ expressed as  

\begin{equation}
P_d(r_i, \theta_i) = P_t Kr_i^{-\eta}G_T(\theta_i)G_R(\phi_i) X_i 
\end{equation}
where $G_R(\phi_i)$ is the antenna gain of the directional receiving antenna and $\phi_i$ represents the angle between the directional receiving antenna and the direction of the transmitting antenna (Fig. \ref{reseauDir} and \ref{dirantenna}).

\subsection{Expression of the SINR} \label{SINRexpression}

The expression (\ref{SINR}) of the SINR can be expressed, for each sub-carrier (dropping the indexes $f$ and $h$): 

\begin{equation} \label{SINRdirect}
\gamma(r_i,\theta_i)=\frac{P_t Kr_i^{-\eta}G_T(\theta_i)G_R(\phi_i) X_i }{
\sum\limits_{j\in\mathcal{N},j\neq i}P_t Kr_j^{-\eta}G_T(\theta_j)G_R(\phi_j) X_j  + N_{th}}.
\end{equation}
We notice that the antenna gain $G_R$ = 1 for omnidirectional receiving antennas.

\begin{figure}[htbp]
\centering
\includegraphics[scale=0.40]{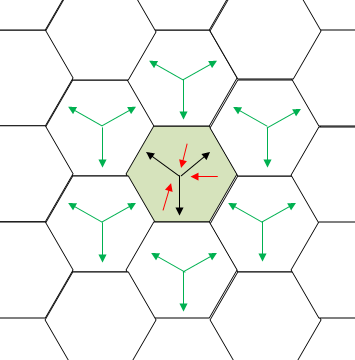}
\caption{\footnotesize Hexagonal network with directional receiving UE (red arrow toward the BS).}
\label{reseauDir}
\end{figure}

\begin{figure}[htbp]
\centering
\includegraphics[scale=0.40]{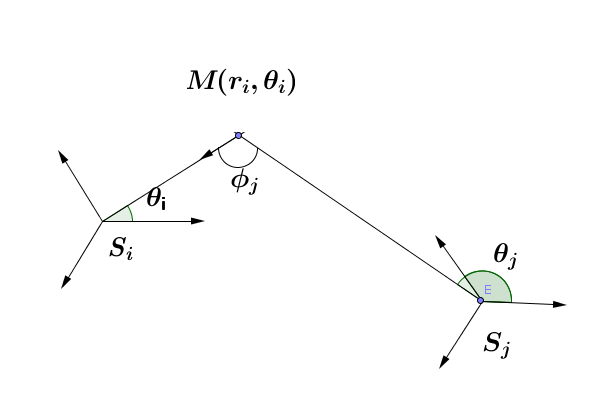}
\caption{\footnotesize User equipment $M$ located at $(r_i, \theta_i)$. It receives a useful power from antenna $i$ and interference power from antenna $j$. The directional antenna of UE is directed toward the antenna $i$}.
\label{dirantenna}
\end{figure}

The angle $\phi_j$ is the angle between the directional receiving antenna of the UE $i$ (red arrows directed toward the BS in the central cell in Fig. \ref{reseauDir}) and the transmitting antenna $j$. So we have $\phi_j = \theta_j  - \theta_i$ (Fig. \ref{dirantenna}) . Therefore, as directional antennas are directed toward the $BS_i$, on the numerator of  (\ref{SINRdirect}) we have $\phi_i$= 0 and $G_R(\phi_i)$= 1 (by using expression (\ref{GRphi}) of the antenna gain $G_R$). We can express (\ref{SINRdirect}) as:

\begin{equation} \label{SINRdirect2}
\gamma(r_i,\theta_i)=\frac{P_t Kr_i^{-\eta}G_T(\theta_i) X_i }{
\sum\limits_{j\in\mathcal{N},j\neq i}P_t Kr_j^{-\eta}G_T(\theta_j)G_R(\phi_j) X_j  + N_{th}}
\end{equation}

In the aim to analyze the specific impact of the use of directional receiving antennas instead of omnidirectional ones, it is interesting to establish the SINR without shadowing.  
Indeed, in this way, it is more easy to propose an interpretation of this impact, since it is not coupled to the influence of a lognormal random variable. 


Considering a density $\rho_{S}$ of sites $S$ and following the approach developed in \cite{KeCoGo07} \cite{KeCoGo10}, let us consider a UE located at $(r_i, \theta_i)$ in the area covered by the BS$_i$. 

The denominator of (\ref{SINRdirect2}) can be expressed as:

\begin{eqnarray} \label{Interference}
I &=& \int P_t \rho_{S} Kr^{-\eta}G_T(\theta)G_R(\theta-\theta_i)r dr d\theta  \nonumber \\ 
&+& \sum_{a=2}^3 G^a_T(\theta_a)G_R(\theta_j-\theta_a)+ N_{th}
\end{eqnarray} 
where the integral represents the interference due to all the other sites of the network, and the discrete sum represents the interference due to the 2 base stations co-localized with the base station $i$. The index $a$ holds for these 2 BS.
Let notice that $\theta_j-\theta_a$ = 0. So we have $G_R(\theta_j-\theta_a)$ = 1. 
Moreover, considering the three sectors of a site $S$, we have for any angle $\theta$\\ 
$G_T(\theta)= G^1_T(\theta)= G^2_T(\theta+2 \pi/3)=G^3_T(\theta-2 \pi/3)$,

Since each site is equipped by 3 antennas we can express (\ref{Interference}) as:  

\begin{eqnarray}  \label{Interference2}
I &=& \int P_t \rho_{S} Kr^{-\eta}G_T(\theta)G_R(\theta-\theta_i)r dr d\theta \nonumber \\ 
&+& P_t K r_i^{-\eta} \sum_{a=2}^3 G^a_T(\theta_a)+ N_{th}
\nonumber \\ 
&=& \int P_t \rho_{S} Kr^{-\eta} r dr \int \sum_{a=1}^3 G^a_T(\theta_i)G_R(\theta-\theta_i)d\theta
\nonumber \\
&+& P_t K r_i^{-\eta} \sum_{a=2}^3 G^a_T(\theta_a)+ N_{th}
\nonumber \\ 
&=& \int P_t \rho_{S} Kr^{-\eta} r dr \times 3 \int G_T(\theta)G_R(\theta-\theta_i)d\theta
\nonumber \\
&+& P_t K r_i^{-\eta} \sum_{a=2}^3 G^a_T(\theta_a)+ N_{th}
\end{eqnarray}

\begin{eqnarray}  \label{Interference2}
I &=& \int P_t \rho_{S} Kr^{-\eta} r dr \int \sum_{a=1}^3 G^a_T(\theta_i)G_R(\theta-\theta_i)d\theta
\nonumber \\
&+& P_t K r_i^{-\eta} \sum_{a=2}^3 G^a_T(\theta_a)+ N_{th}
\nonumber \\ 
&=& \int P_t \rho_{S} Kr^{-\eta} r dr \times 3 \int G_T(\theta)G_R(\theta-\theta_i)d\theta
\nonumber \\
&+& P_t K r_i^{-\eta} \sum_{a=2}^3 G^a_T(\theta_a)+ N_{th}
\end{eqnarray}

%

%

The approach developed in \cite{KeCoGo07} \cite{KeCoGo10} allows to express $\int P_t \rho_{S} Kr^{-\eta} r dr$  as $\frac{2\pi\rho_{S}P_tK(2R_c-r)^{2-\eta}}{\eta-2}$, where $2 R_c$ represents the intersite distance.
We refer the reader to \cite{KeCoGo07} \cite{KeCoGo10} for the detailed explanation and validation through Monte Carlo simulations. 
Therefore, (\ref{Interference2}) can be expressed:
\begin{eqnarray}  \label{interferencefluid}
I &=& \frac{6 \pi P_t (2R_c-r)^{2-\eta}}{\eta-2} \rho_{S} K \int_0^{2 \pi} G_T(\theta)G_R(\theta-\theta_i)d\theta \nonumber \\ 
&+& P_t K r_i^{-\eta}\sum_{a=2}^3 G^a_T(\theta_a)+ N_{th}
\end{eqnarray}  

The SINR (\ref{SINRdirect2}) $\gamma(r_i,\theta_i)$ is finally given by the expression: 
\begin{eqnarray}  \label{SINRfluid2}
\frac{1}{\gamma(r_i,\theta_i)}&=& \frac{6 \pi \rho_{S} (2R_c-r_i)^{2-\eta}}{(\eta-2)r_i^{-\eta}} \frac{\int_0^{2 \pi} G_T(\theta)G_R(\theta-\theta_i)d\theta}{G_T(\theta_i)} \nonumber \\ 
&+& \frac{\sum_{a=2}^3 G^a_T(\theta_a)}{G_T(\theta_i)}+ \frac{N_{th}}{P_t Kr_i^{-\eta}G_T(\theta_i)}
\end{eqnarray}  

%

%
%

\subsection{Interest of the analytical formula}
The classical way to compute SINR given by  (\ref{SINRdirect}) needs to take into account all the distances between the UE and the base stations. Moreover, the formula is intractable. Therefore, it is needed to approximate or to do simulations, even if the shadowing is not taken into account. 

Conversely, the formula (\ref{SINRfluid2}) expresses the SINR by only considering the distance between the UE and the serving BS. This formula is tractable and a simple numerical calculation is needed to calculate it.
Another advantage of this formula is that it focuses on the important parameters of the system, 
such as intersite distance, propagation parameter, transmitting antenna gain. It also highlights on the impact of directional antennas, through the receiving antenna gain, and shows how their use can decrease the interferences and consequently  increase the SINR. 


\subsection{Throughput calculation}
The SINR allows to calculate the reachable throughput $D_u$ of an UE $u$, by using Shannon expression. 
For a bandwidth $W$, it can be written:
\begin{equation} \label{Dugamma}
D_u = W \log_2(1+\gamma_u)
\end{equation}
Expression (\ref{Dugamma}) enables to calculate the theoretical maximum achievable throughputs. \\

\textbf{Remarks:} Let notice that the mapping between the received SINR and the achieved throughput are established by the mean of \textit{link curves} in the case of realistic network systems.\\

\section{Scenarios and Assumptions} \label{scenario}

\subsection{Scenarios} \label{scenario1}

Our aim consists in analyzing the interest to deploy UE equipped with Directional Receiving Antennas instead of Omnidirectional Receiving Antennas. Various parameters may have an impact on the SINR, the throughput and the coverage of BSs. We present hereinafter the parameters we chose in our analysis.
We analyze different scenarios corresponding to the situations which may happen in a real network:
\begin{itemize}
\item sub-urban environment: ISD = 2000m
\item rural environment: ISD = 5000m
\item rural environment: ISD = 10000m
\end{itemize}

For each scenario, we consider two kinds of directional receiving antennas:

\begin{itemize}
\item aperture of $35^\circ$
\item aperture of $17.5^\circ$
\end{itemize}

\subsection{Assumptions} \label{assumptions}

Let us consider:
\begin{itemize}
\item Hexagonal network composed of sectored sites
\item Three base stations per site
\item Antenna gain of transmitting BS is given by (in dB)
\begin{equation} \label{GTheta}
G_T(\theta) = - \min\left[12\left(\frac{\theta}{\theta_{3dB}}\right)^2, A_m \right],
\end{equation}
where $\theta_{3dB}$ = $70^\circ$ and $A_m$ =25 dB
\item Antenna gain of directional receiving antennas is given by (in dB)
\begin{equation} \label{GRphi}
G_R(\phi) = - \min\left[12\left(\frac{\phi}{\phi_{3dB}}\right)^2, A_m \right],
\end{equation}
where ($\phi_{3dB}$,$A_m$) = ($35^\circ$, 23 dB) or ($17.5^\circ$, 21 dB)

\item
downlink OFDM LTE, carrier frequency 2.6 GHz, channel bandwidth 10MHz,
\item 
the transmitting power: we set it at 46 dBm, as in a realistic transmission environment.
\item 
standard deviation of the shadowing $\sigma= 8$dB.
\end{itemize}

\section{Results} \label{Resultats}

The analysis is focused on the analysis of the quality of service, performance and coverage. 
The establishement of the cumulative distributed functions (CDF) of the SINR represents an important characteristic of the system. Indeed, they first allow to characterize the coverage and the outage probability. They also characterize the performance distribution, and the quality of service that can be reached by the system.

%
%
%
\begin{figure}[htbp]
\centering
\includegraphics[scale=0.35]{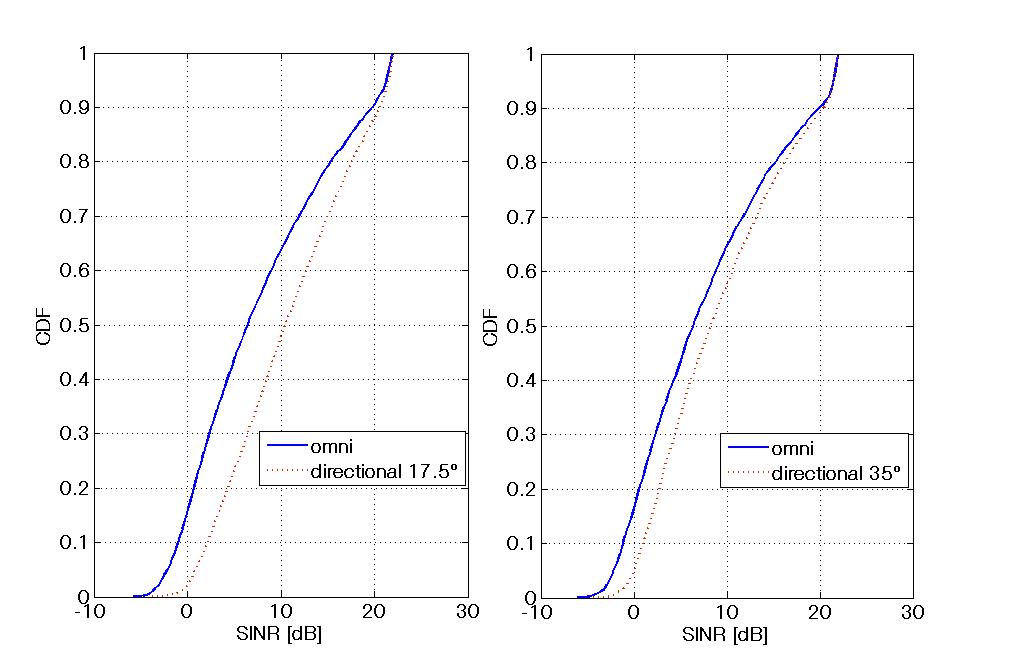}
\caption{\footnotesize CDF of the SINR for suburban environment (ISD= 2000m), ominidirectional receiver compared to directional one with aperture $17.5^{\circ}$ (left) and $35^{\circ}$ (right)}
\label{CDF2000}
\end{figure}

\begin{figure}[htbp]
\centering
\includegraphics[scale=0.35]{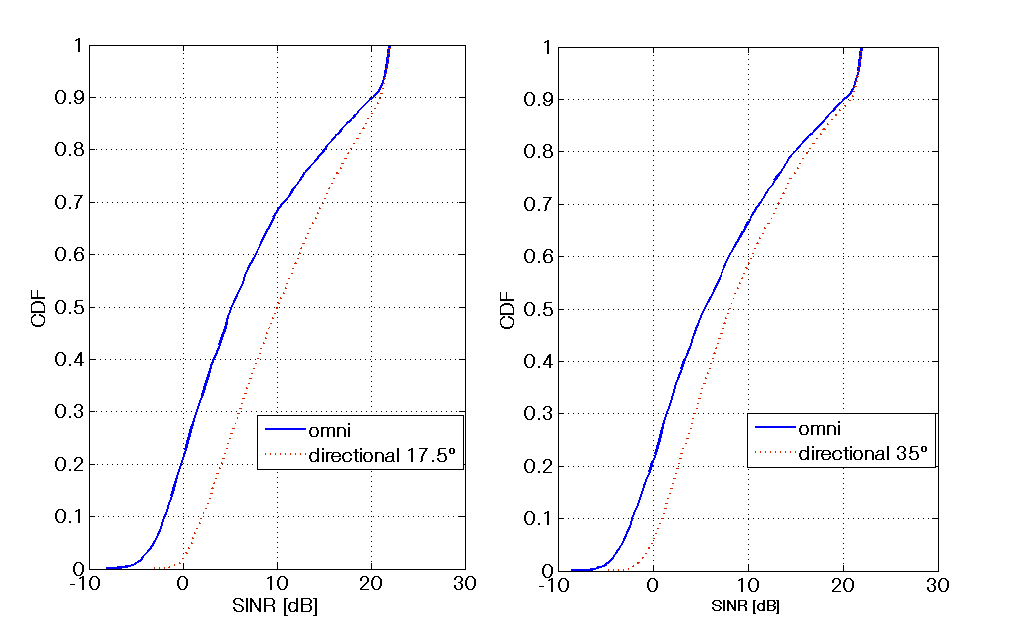}
\caption{\footnotesize CDF of the SINR for rural environment (ISD= 5000m), ominidirectional receiver compared to directional one with aperture $17.5^{\circ}$ (left) and $35^{\circ}$ (right)}
\label{CDF5000}
\end{figure}

\begin{figure}[htbp]
\centering
\includegraphics[scale=0.35]{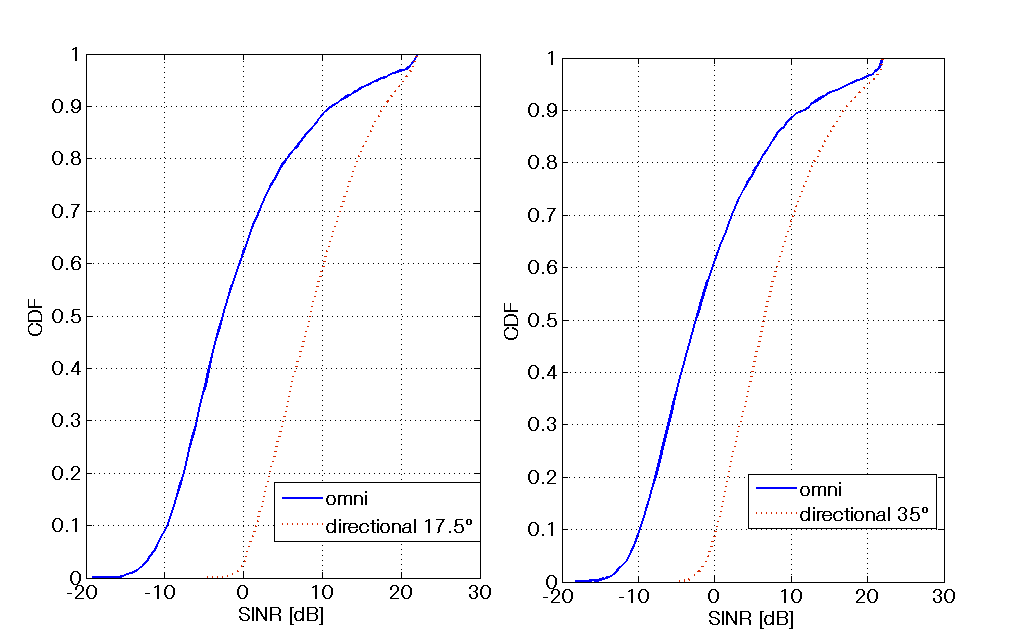}
\caption{\footnotesize CDF of the SINR for rural environment (ISD= 10000m), ominidirectional receiver compared to directional one with aperture $17.5^{\circ}$ (left) and $35^{\circ}$ (right)}
\label{CDF10000}
\end{figure}

\subsection{Impact of the intersite distance on SINR distribution}\label{ISDimpact}

This analysis consists in determining  the interest to deploy directional antennas according to the kind of environment: rural or semi-urban. 
Figures \ref{CDF2000}, \ref{CDF5000} and \ref{CDF10000} show the cumulative density function (CDF) of the SINR offered to the UE, in all scenarios considered.
These curves show that, when the intersite distance increases, the SINR reached by UEs equipped with omnidirectional antennas decrease. For example, considering an outage probability of 0.1, the decrease of the SINR may reach 9~dB. Indeed, for an ISD of 2000 m the SINR reaches -1~dB (Fig. \ref{CDF2000} left) and for an ISD of 10000 m the outage probability decreases until -10~dB (Fig. \ref{CDF10000} left). There is also a decrease for UEs equipped by directional receiving antennas. However, this decrease is very low. In particular, it only reaches 0.7 dB in the same case as before.
This result can be explained as follows. The interference due to the other base stations of the network is larger by using omnidirectional antennas than by using directional ones (cf. (\ref{SINRfluid2})). When the ISD increases, the impact of thermal noise becomes more and more important. However, the relative impact is much larger in the case of omnidirectional antennas than in the case of directional antennas.
We can thus conclude that 

(i) whatever the scenario considered, a deployment of UE equipped with directional antennas allows to improve the CDF of the SINR (compared to the case where UE are equipped with omnidirectional antennas). Therefore the outage probability decreases,  and the performance  and QoS increase,

(ii) since the coverage of a BS is an increasing function of the SINR, the use of directional antennas improves the coverage of a BS, too. 

(iii) the performance and QoS of UE equipped by directional antennas are \textit{much less sensitive} to the intersite distance, and thus to the type of environment (rural or semi-urban).

\subsection{Impact of the aperture angle on SINR distribution}\label{Apertureimpact}
Let us consider the same curves as before. Figures \ref{CDF2000} and \ref{CDF10000} show that an aperture of $17.5^\circ$ allows to reach higher SINR than an aperture of $35^\circ$. For example, considering an outage probability of 0.1, the decrease of the SINR reaches 1 dB by the use of an antenna with aperture of $35^\circ$ instead of $17.5^\circ$ (Fig. \ref{CDF2000} left compared to right with ISD=2000 m). This decrease reaches about 1.5 dB for an ISD of 10000 m (Fig. \ref{CDF10000} left compared to right). 
Therefore, a low aperture allows a higher improve of the SINR than a wide one. It is due to the expression of the interference (\ref{interferencefluid}) which depends on $G_R$. With a low aperture, the impact of the interference due to the other base stations of the network decreases. 
Let us however notice that the maximum difference observed on these curves reaches 2 dB. It is relatively low.

\subsection{Impact on the SINR of each UE}\label{UEimpact}
The CDF observed in Fig. \ref{CDF2000}, \ref{CDF5000} and \ref{CDF10000} show that the use of directional receiving antennas improves the quality of service and the performance, and that it allows to decrease the outage probability in all cases. 
We could conclude that  it is interesting, in all cases, to use a directional receiving antenna than an omnidirectional one. 
However, these curves do not focus of the impact of the use of directional antennas \textit{on each point of the system}.
Therefore, it seems interesting to analyze this point of view:

(i) by calculating for each UE, the difference between the SINR reached when this UE is equipped by a directional antenna, and the SINR reached if it is equipped by an omnidirectional one, 

(ii) and by drawing the CDF of that difference of SINR. 

\begin{figure}[htbp]
\centering
\includegraphics[scale=0.45]{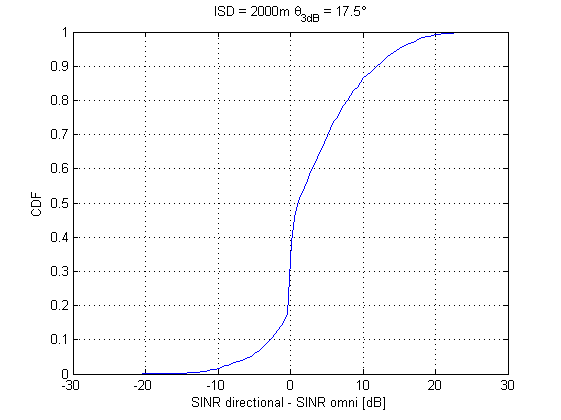}
\caption{\footnotesize CDF of the difference at each point of the SINR 'directional' (aperture $17.5^{\circ}$) and the SINR 'omnidirectional' for semi-urban environment (ISD= 2000m)}
\label{cdfdiff2000}
\end{figure}

This curve is represented in Fig. \ref{cdfdiff2000}.
It can be observed a very interesting result: that difference may be positiv, negativ, or null. 

This result means that for some UEs of the system, it is \textit{not interesting} to replace omnidirectional receivers by directional ones. 
It seems to be in contradiction with the other observations (Fig. \ref{CDF2000}, \ref{CDF5000} and \ref{CDF10000}), which showed that the use of directional antennas improved the system in all cases.
In fact, these last curves are the expression of the global behaviour of the whole system. They do not provide information for each individual UE. 
It means that, though globally increase by using directional antennas, the SINR may \textit{locally} decrease.
Fig. \ref{cdfdiff2000} also shows that, in the 2000 m ISD case, for about 20\% of the UE, the use of directional antennas decrease the SINR (0 $\leq$  CDF $\leq$ 0.2). For a proportion of 20\% of the UE the use of directional antennas neither increase nor degrade the SINR (0.2 $\leq$  CDF $\leq$ 0.4) . And for 60\% directional antennas increase the SINR.

Another interesting result of this curve consists in the observation of the amplitude of the improvement. Indeed, Fig. \ref{CDF2000} shows that the difference between omni directional SINR and directional one reaches a maximum of about 5 dB. In contrast, Fig. \ref{cdfdiff2000} shows that that difference may reach 20~dB! This result can be interpreted as follows. 
The UEs which have low SINR by using omnidirectional antennas may have a large improve of SINR by using directional antennas. 
However, the use of directional antennas may degrade the SINR of some UEs, and this degradation may reach 15 dB.

\subsection{Impact of the shadowing on SINR distribution}\label{Shadowingimpact}
\begin{figure}[htbp]
\centering
\includegraphics[scale=0.35]{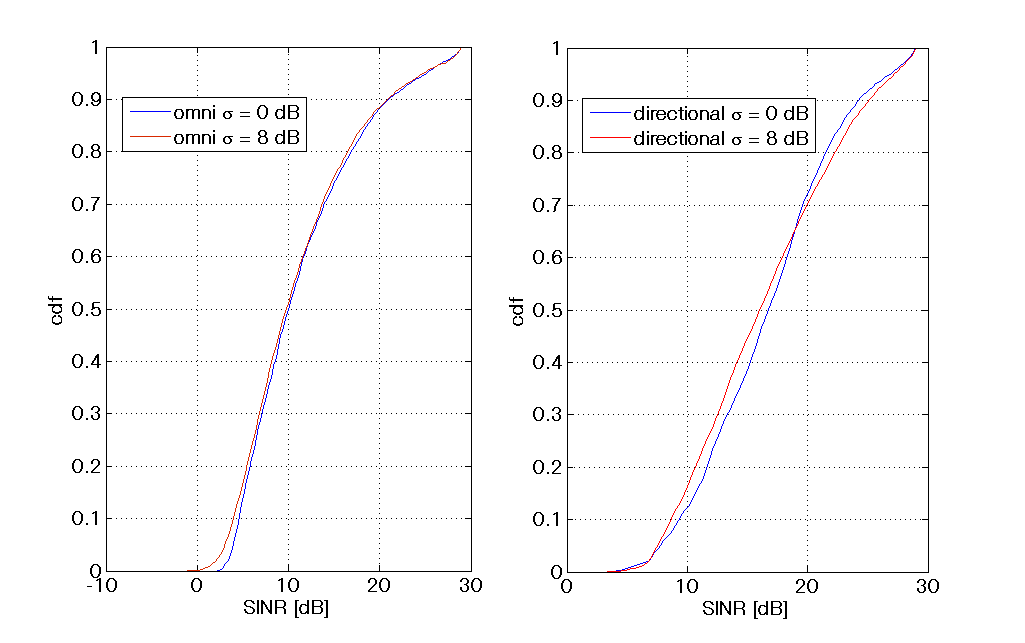}
\caption{\footnotesize CDF of the SINR for suburban environment (ISD= 2000m), ominidirectional receiver compared to directional one with aperture $17.5^{\circ}$ (left) and $35^{\circ}$ (right)}
\label{CDFSINR_OmniDirSigma0-8dB_5000-17}
\end{figure}

Since we established a formula of the SINR without considering the shadowing, it appears interesting to analyze the impact of the shadowing, too. In this aim, we compare simulations by considering shadowing and simulations without shadowing. Let us remind that UE are connected to the BS which offers the highest useful power. Fig.\ref{CDFSINR_OmniDirSigma0-8dB_5000-17} shows that the impact of shadowing on the distribution of the SINR is relatively low. 
It can be observed that for omnidirectional antennas, the distributions with and without shadowing are identical except for low values of SINR: the difference reaches about 1 dB for an outage of 2\%.
In the case of directional antennas, the curves are very close: the maximum difference observed is about 1 dB. 
Therefore, it seems justified to develop an analysis model without considering the shadowing. Moreover that analysis allows to establish a simple analytical expression of the SINR. This one allows to establish performance and quality of service in a simple way. 

\emph{Remark} : This result is due to the fact that UE are connected to their best serving station, i.e. the BS which offers the highest useful signal.

\section{Conclusion} \label{conclude}
We propose an approach, based on the use of directional receiving antennas on UEs, to mitigate interferences in wireless networks. We show that solution allows an improvement of the CDF of the SINR, therefore an improvement of performance and quality of service, and we quantify it. 
Although the SINR is degraded for some UEs, most of them have a high improvement of their SINR. 
Moreover, we establish that this solution is very few sensitive to the environment, rural or semi-urban. And it is less complex to implement and deploy than other solutions based on ICIC or CoMP. 
This solution seems to perfectly fit with high data rate Internet access.


\begin{thebibliography}{1}


\bibitem{Lee12}
Daewon Lee, Geoffrey Y. Li and Suwen Tang
Inter-Cell Interference Coordination
for LTE Systems, Globecom 2012

\bibitem{Kos13}
Chrysovalantis Kosta, Bernard Hunt,Atta U. Quddus, Rahim Tafazolli
Improved Inter-cell Interference Coordination
(ICIC) for OFDMA multi-cell systems,
European Wireless 2013


\bibitem{iichetnet}
D. Lopez-Perez, I. Guvenc, G. De La Roche, M. Kountouris, T. Q. S. Quek and J. Zhang, Enhanced Intercell Interference Coordination Challenges in Heterogeneous Networks, IEEE Wireless Communications, June 2011. 



\bibitem{Zha12}
Xinyu Zhang, Mohammad A. Khojastepour, Karthikeyan Sundaresan,
Sampath Rangarajan, Kang G. Shin
Exploiting Interference Locality in Coordinated
Multi-Point Transmission Systems,
ICC 2012

\bibitem{Ngu11}
L.-H. Nguyen, R. Rheinschmitt, T. Wild, S. ten Brink
Limits of Channel Estimation and Signal
Combining for Multipoint Cellular Radio (CoMP),
International Symposium on Wireless Communication Systems,  2011


\bibitem{Che11}
Dorra Ben Cheikh Battikh, Jean-Marc Kelif, Marceau Coupechoux and Philippe Godlewski
Dynamic System Performance of SISO, MISO and
MIMO Alamouti Schemes,
Sarnoff Symposium, 2011

\bibitem{Sch11}
Stefan Schwarz, Michal Simko and Markus Rupp,
On Performance Bounds for MIMO OFDM Based Wireless Communication Systems,
International Workshop on Signal Processing Advances in Wireless Communications, 2011


\bibitem{Kai00}
S. Kaiser, Spatial Transmit Diversity Techniques for Broadband OFDM Systems, Proc. of Globecom, 2000. 

%
%
%
%



\bibitem{KeCoGo07}
J.-M. Kelif, M. Coupechoux and P. Godlewski, Spatial Outage Probability for Cellular Networks, Proc. of GLOBECOM, 2007. 


\bibitem{ITUR2009}
Report ITU-R M.2135-1, Guidelines for evaluation of radio interface
technologies for IMT-Advanced, 12/2009


\bibitem{3GPP209}
3GPP TSG-RAN1 WG1, LTE Downlink Performance, Conference Call, Apr 24th 2007. R1-071978.

%

\bibitem{KeCoGo10}
J-M. Kelif, M. Coupechoux and P. Godlewski, On the Dimensioning of Cellular OFDMA Networks, Physical Communication Journal, Ref : PHYCOM118, online October 2011, DOI : 10.1016/j.phycom.2011.09.008.




\end{thebibliography}
\end{document}